\documentclass[12pt,preprint]{aastex}

\shorttitle{GQ Lup B Photometric Analysis}

\shortauthors{Marois et al.}

\begin{document}

\title{GQ Lup B Visible \& Near-Infrared Photometric Analysis}

\author{Christian Marois\altaffilmark{a}, Bruce Macintosh\altaffilmark{a}, Travis Barman\altaffilmark{b}}
\altaffiltext{a}{Institute of Geophysics and Planetary Physics L-413, Lawrence Livermore National Laboratory, 7000 East Ave, Livermore, CA 94550}
\altaffiltext{b}{Lowell Observatory, 1400 W. Mars Hill Rd., Flagstaff, AZ 86001}

\begin{abstract}
We have re-analyzed archival HST and Subaru data of the recently discovered planetary mass companion (PMC) GQ Lup B. With these we produce the first R and I band photometry of the companion and fit a radius and effective temperature using detailed model atmospheres. We find an effective temperature of $2335 \pm 100$K, and a radius of $0.38 \pm 0.05$R$_{\odot}$ and luminosity of $\log(L/L_{\odot}) = -2.42 \pm 0.07$ (at 140pc). Since we fit wavelengths that span most of the emitted radiation from GQ Lup this luminosity estimate is robust, with uncertainty dominated by the distance uncertainty ($\pm 50$pc). The radius obtained for 140pc (0.38R$_{\odot}$) is significantly larger than the one originally derived and larger than model predictions. The mass of the object is much more model-dependent than the radiative properties, but for the GAIA dusty models we find a mass between 10-20 M$_{\rm{Jup}}$, in the range of the brown dwarf and PMC deuterium burning boundary. Assuming a distance of 140pc, observations fit to 1$\sigma$ the Baraffe evolution model for a $\sim 15$ M$_{\rm{Jup}}$ brown dwarf. Additionally, the F606W photometric band is significantly overluminous compared to model predictions and other brown dwarfs. Such overluminosity could be explained by a bright H$\alpha$ emission from chromospheric activity, interaction with another undetected companion, or accretion. Assuming that GQ Lup B has a bright H$\alpha$ emission line, its H$\alpha$ emission strength is $10^{-1.71 \pm 0.10}L_{\rm{bol}}$, significantly larger than field late-type dwarfs. GQ Lup B might be strongly accreting and still be in its formation phase.
\end{abstract} 
\noindent{\em Suggested running page header:} GQ Lup B Photometric Analysis

\section{Introduction}
Direct exoplanet detection around stars is a challenging endeavor, but possible for companions to the youngest stars \citep{neuhauser2005,chauvin2005a,chauvin2005b}. Unlike the exoplanets found by precision radial velocity techniques that provide little information (albeit model independent) about the physical properties of planetary mass companions (PMCs), the physical properties of these PMCs must be inferred by comparing atmosphere and evolution models to observed spectra or photometry.

One of the most recent PMC candidates identified orbits the star GQ Lup \citep{neuhauser2005} (N05). This star is a young K7eV TTauri star in the Lupus I cloud \citep{tachihara1996}. The star has an estimated age of less than 2Myr (N05) and is situated at 140pc with a potential range from 90 to 190pc \citep{wichmann1998,neuhauser1998,knude1998}. The original analysis of the GQ Lup PMC candidate, called GQ Lup B, is based on K- and L$^{\prime}$-band photometry along with a K-band spectrum. From this early work, an effective temperature of $2050 \pm 450$K, a radius of $0.12 \pm 0.06$R$_{\rm{\odot}}$, a luminosity of $\log(L/L_{\odot}) = -2.37 \pm 0.41$ and $\log g$ of $2.52 \pm 0.77$ are obtained \citep{neuhauser2005,neuh2005b}.

Since other wavelength bands are available from HST and Subaru that span most of the radiation emitted by GQ Lup B, it is possible to conduct a more complete photometric analysis. This analysis is presented below along with the implications of a detected R-band overluminosity.

\section{GQ Lup Photometry}\label{phot}
In addition to the (N05) VLT data, GQ Lup B has been previously observed by both the Subaru telescope, program o02312, and the Hubble Space Telescope (HST), programs SNAP7387 and 9845. The coronagraphic imager with adaptive optics instrument (CIAO) \citep{murakawa2004} was used at the Subaru telescope while the Wide Field Planetary Camera No. 2 (WFPC2) and NICMOS were used with HST.

\subsection{HST Visible and NIR Photometry \label{hstphot}}
Data were retrieved from the public MAST Hubble Space Telescope archive at STScI for the filters F606W, F814W, F171M, F190N and F215N using the automated reduction pipeline. The companion is clearly visible 0.7$^{\prime \prime}$ West of GQ Lup in all filters.  The GQ Lup PSF in each filter was first subtracted using reference PSFs of a second star observed in the same program or using simulated PSFs produced by the Tiny Tim software \citep{krist1993}, selecting the one that gives the smallest residual at 0.7$^{\prime \prime}$ separation. The simulated PSFs in each filter were then used to estimate the GQ Lup B flux. PSFs were simulated with five times the sampling, shifted, and binned to the detector resolution. The set of parameters, i.e. fractional pixel PSF position and flux normalization, that minimize the RMS noise inside a $6 \times 6$ pixels box centered on GQ Lup B was kept. Magnitude errors were estimated by calculating the RMS value of the total GQ Lup subtracted PSF residual flux at the same angular separation using the same $6 \times 6$ pixels box but at a different field angle. Regions contaminated by residual flux from the diffraction spider were avoided. Total companion fluxes were obtained by integrating the Tiny Tim simulated PSFs that best subtract the GQ Lup B PSF. The WFPC2 charge-Transfer efficiency bias was corrected \citep{whitmore1999}, but the amplitude of the effect is small, less than 5\%, since GQ Lup is bright (more than 3000 counts inside a 2 pixel radius aperture). Table~\ref{table1} shows the obtained apparent magnitudes and estimated error bars. Interstellar extinction, assuming A$_{\rm{v}} = 0.4 \pm 0.2$ \citep{batalha2001} and the extinction law of \citet{rieke1985}, is also tabulated.

\subsection{Subaru Photometry}
\begin{figure}
\plotone{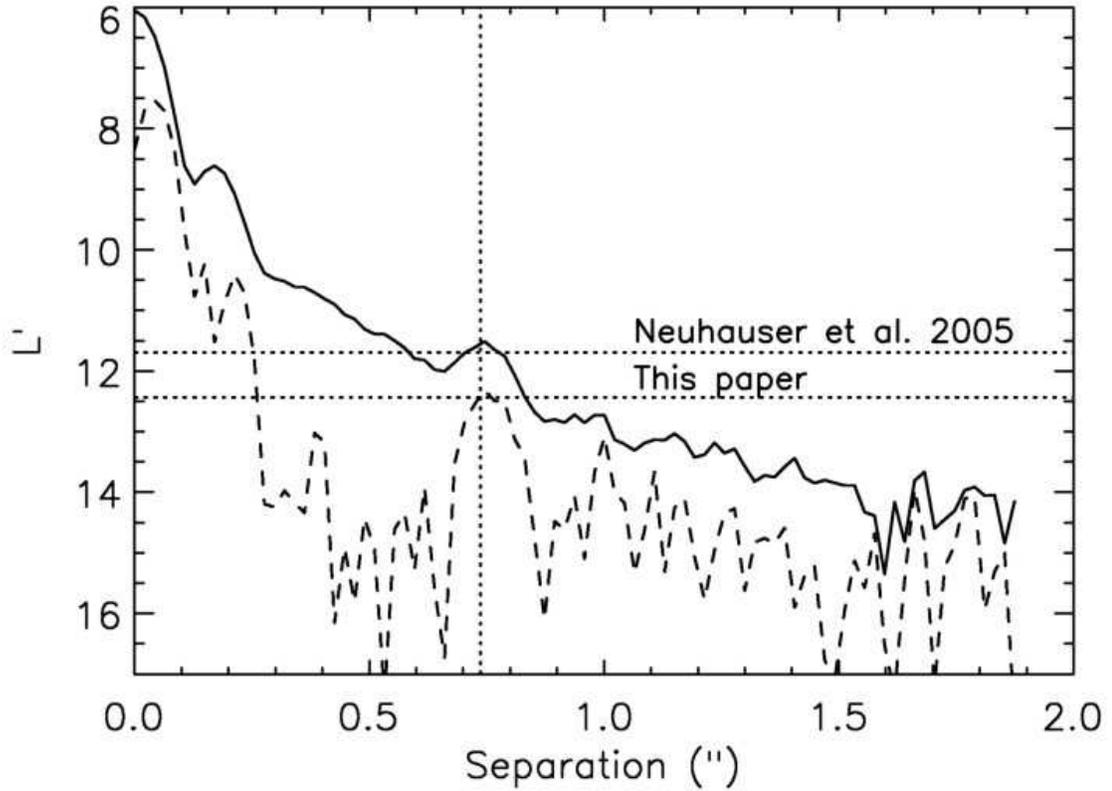}
\caption{GQ Lup B L$^{\prime}$ magnitude measurement. Solid line shows GQ Lup PSF intensity profile going through GQ Lup B. Dotted vertical line shows the separation of GQ Lup B. Dashed line shows the same intensity profile after sky subtraction and GQ Lup PSF halo subtraction. The two horizonthal dotted lines show N05 and our magnitude estimates.\label{fig1}}
\end{figure}

Using the public Subaru archive (SMOKA), data for K, CH$_4$ and L$^{\prime}$ filters were retrieved.  For K and CH$_4$, since a dithered pattern was used to acquire GQ Lup, a sky image was constructed from the median of all acquired images in each band. For L$^{\prime}$, the sky image sequence was used to subtract the thermal background. Since no flat field images are available, images were simply registered at the image center, median combined, and a 180 degrees rotation was used to subtract the smooth PSF halo and residual sky background (an angle of 170 degrees was used for CH$_4$ data due to a bad pixel located at 180 degrees of GQ Lup B position). GQ Lup B is clearly visible in all three filters and was not saturated or occulted behind a coronagraph. The PSF average peak flux per readout in raw images for the CH$_4$, K and L$^{\prime}$ bandpasses are respectively 30,000, 40,000 and 50,000 electrons. PSF peaks are thus below the 60,000 electrons level (CIAO Aladdin-2 detector has a 5\% non-linearity at 60,000 electrons) in all bandpass images, detector non-linearity is thus neglected. Therefore, $\Delta$mag measurements were possible in each filter. GQ Lup B magnitude differences were found by optimizing the PSF subtraction from the PSF of GQ Lup. Magnitude errors were estimated by calculating the flux RMS variation of 10 square boxes of 1.5 $\lambda/D$ width at 10 different angular positions and at the same separation as GQ Lup B. Apparent magnitudes were determined using known H (to determine the CH$_4$ magnitude), K and L$^{\prime}$ magnitudes for GQ Lup, H=$7.70\pm 0.03$~mag, K=$7.10\pm0.02$~mag (2MASS) and L=$6.05\pm0.13$ \citep{glass1974,hughes1994}. Magnitude differences between the L$^{\prime}$ and the L filters are negligible. Our L$^{\prime}$ magnitude differs by 0.74 magnitude compared to the one published by N05 while the K (Ks for N05) magnitudes are consistent to 1~$\sigma$ accuracy (see Table~\ref{table1}). A slice of the L$^{\prime}$ image through GQ Lup A and B is given in Fig.~\ref{fig1}; the N05 measurement is likely contaminated by flux from the primary.

\section{GQ Lup B Temperature and Radius Fit}
\begin{figure}
\plotone{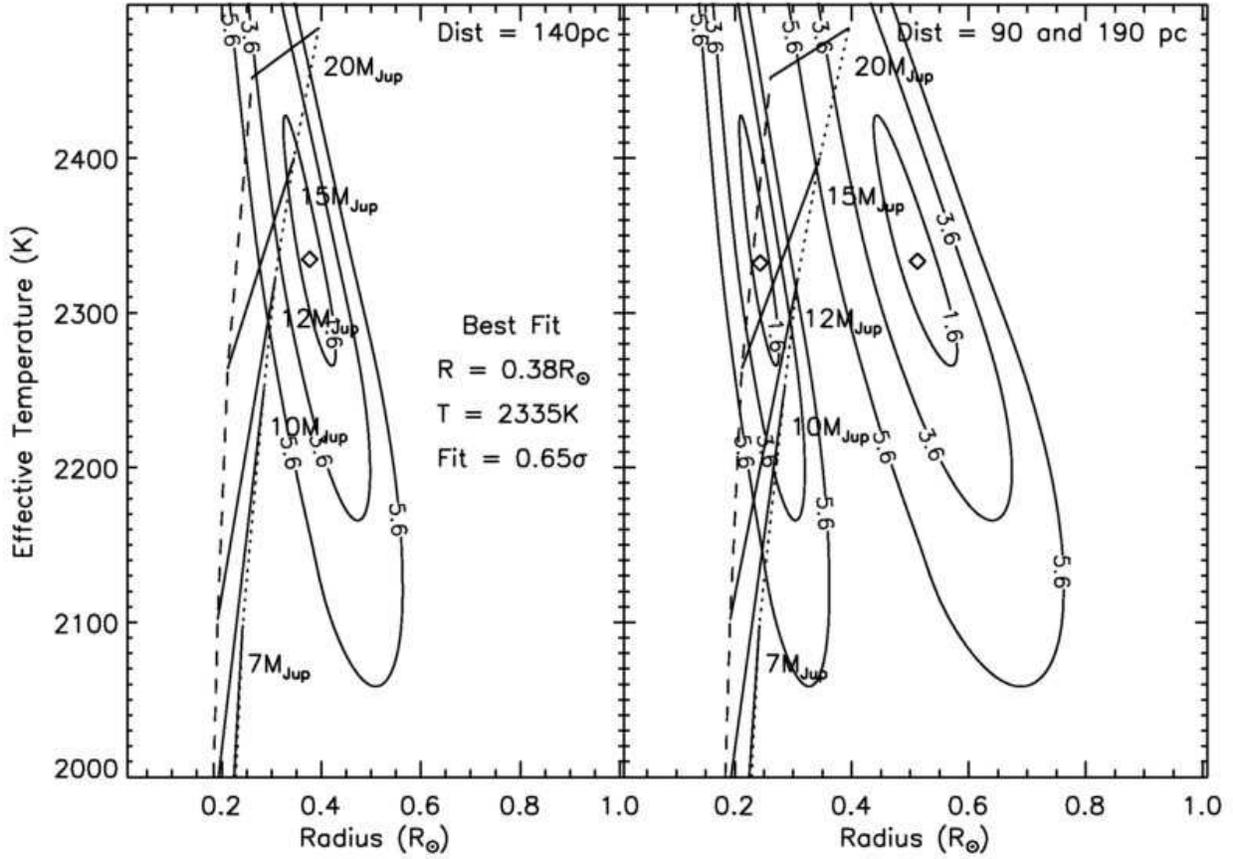}
\caption{Temperature-radius fit for 140 (left) and 90 and 190pc (right) distance. The diamond symbol represents the best fit. Three contour levels at 1, 3 and 5$\sigma$ from the best fit are shown. Model predictions for 7, 10, 12, 15 and 20~M$_{\rm{Jup}}$ and 1 (dotted line) and 5~Myr (dashed line) are also shown.\label{fig2}}
\end{figure}

\begin{figure}
\plotone{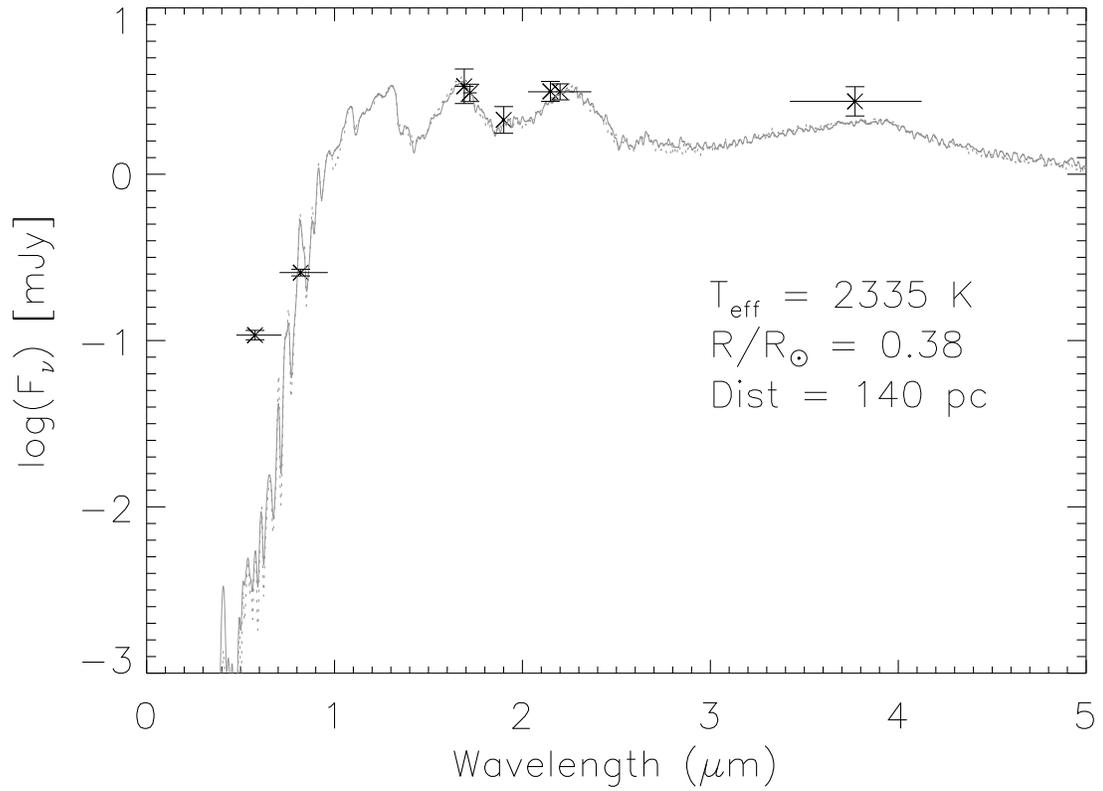}
\caption{Best temperature-radius fit (solid line) for 140pc, $\log g = 3$, typical 0.4 IE in V band and 1 Myr. Luminosity is $\log(L/L_{\odot}) = -2.42$. Dashed line is the same fit with $\log g = 4$.\label{fig3}}
\end{figure}

A solar metalicity subset of the GAIA dusty model atmosphere grid (Hauschildt et al., in preparation) was used to fit the photometry in Sect.~\ref{phot} and determine the radius and temperature of GQ Lup B. The fact the all the data was not acquired simultaneously but spread over many years could suggest a problem related to a potential GQ Lub B (and A for contrast ratio measurements) variability. Since the HST, VLT and Subaru H and K photometry were acquired two years apart and agree to $\sim 1\sigma$ suggest that GQ Lup B (and A) variability is small for our dataset. Given that surface gravity has only a small effect on predicted broad band photometry and that such young low-mass objects typically have low gravities, a standard $\log g = 3$ was assumed. This gravity is consistent with the one found by N05 from a low resolution K-band spectrum, i.e $\log g = 2.52 \pm 0.77$.  The model fluxes were convolved and integrated over the appropriate transmission curves for each filter listed in Table~1. For the F814W HST filter, the total optical system transmission was used instead of the filter transmission due a drop of sensitivity in a spectral region where GQ Lup B shows a significant increase of luminosity. These synthetic broad-band flux densities were interpolated to produce a uniform square grid in radius and temperature space.  Observed magnitudes were corrected for distance and estimated interstellar reddening and transformed to fluxes using a calibrated Vega spectrum.  The shape of the observed broad-band SED from $R$ to $L$-band determines the effective temperature while the overall scaling needed to best match the observed fluxes (for a fix distance) determines the radius.  A standard deviation for each model, $\sigma_{\rm{model}}$, was calculated using the following standard equation,
 
\begin{equation} 
\sigma_{\rm{model}} = \sqrt{\frac{1}{N-1} \sum_{i=1}^{N} \left( \frac{o_i-m_i}{e_i}\right) ^2},
\end{equation}

\noindent where $N$ is the number of photometric measurements and $o$, $e$ and $m$ are respectively the observe data point, its error bar, and expected model-derived flux. The F606W bandpass was not included since this point is clearly overluminous compared to the near-infrared bands (see Sect.~\ref{disc}). Figure~\ref{fig2} shows the temperature as a function of radius for GQ Lup B. Error contours are defined as models that deviate by 1, 3 and 5$\sigma$ from the best fit model. The predicted temperature and radius of the \citet{baraffe2003} evolution models for 1 and 5~Myrs and 7, 10, 12, 15 and 20~M$_{\rm{Jup}}$ are also shown. The best fit found is at 0.65$\sigma$ from the model with an effective temperature of $2335$K and radius of $0.38$R$_{\odot}$ (assuming a distance of 140pc). The fit was then rerun for both maximum and minimum allowed distance and interstellar extinction to estimate error bars (see Table~\ref{table2}). The overall fit to account for both the distance and interstellar extinction is thus an effective temperature of $2335\pm 100$K and a radius of range of $0.38 \pm 0.05 d_{140}$ R$_{\odot}$, where $d_{140} = d/140$ and $d$ is the distance in parsecs (see Fig.~\ref{fig3}). Note that the fit would be equally consistent with a $\log g = 4$ in agreement with the fact that broadband photometry is insensitive to $\log g$ (the mass of GQ Lup B can thus not be contrained by our adopted $\log g$ and derived radius). The resulting luminosity, accounting for the correlated error bars, is equal to $\log(L/L_{\odot}) = -2.42 \pm 0.07 + \log(d_{140}^2)$. Interstellar extinction does not significantly change the derived effective temperature but the distance adds a possible range three times the derived radius error bar. A better distance estimate would be required to further constrain the radius and the luminosity of GQ Lup B. If the GQ Lup system is really at 140pc, for our best temperature fit, our radius measurement is larger (by 3$\sigma$) than model predictions (0.32 R$_{\odot}$), a result similar with what is observed in young star clusters \citep{mohanty2004}. The larger than expected radius could also be explained if GQ Lup B is actually an unresolved binary. At a 1$\sigma$ confidence level, our derived effective temperature is consistent but has an error bar five times smaller than the one derived by N05 and our radius is more than three times larger. Note that the radius ($0.12 \pm 0.06$R$_{\rm{\odot}}$), effective temperature ($2050 \pm 450$) and luminosity ($\log(L/L_{\odot}) = -2.37 \pm 0.41$) given in \citet{neuhauser2005,neuh2005b,guenther2005} are incompatible; for such radius and temperature, the luminosity should be closer to $\log(L/L_{\odot}) = -3.67$, clearly inconsistent with the observed photometry (see Fig.~\ref{fig3}). It is clear from Fig.~\ref{fig3} that the F606W magnitude is overluminous compared to model predictions -- the implications of this are discussed below. GQ Lup B is also overluminous in the R-band when compared to other brown dwarfs (see, for example, the visible spectra of Pleiades brown dwarfs \citep{martin2000}).

As discussed in N05, planetary evolution models are highly uncertain for young objects. However, it is still interesting to compare our results to the \citet{baraffe2003} evolution models, even if we know that initial conditions introduce uncertanity for $\sim $1Myr objects \citep{baraffe2002}. Such problems have been confirmed by observations \citep{koenig2002,mohanty2004,reiners2005,stassun2006}. For the acceptable distance range (90 to 140pc) and age ($< 2$~Myr), our temperature and radius are consistent with a 10-20~M$_{\rm{Jup}}$ PMC/brown dwarf at 1$\sigma$, at the boundary between a PMC and brown dwarf similar to the AB Pic PMC candidate \citep{chauvin2005b}. Assuming a distance of 140pc, a slightly higher effective temperature of 2400K and a smaller radius, our observations are consistent with a $\sim 15$ M$_{\rm{Jup}}$ brown dwarf (Baraffe models) to 1$\sigma$ accuracy.

\section{Discussion}\label{disc}
The HST F606W magnitude is significantly overluminous compared to model predictions (three magnitudes, see Fig~\ref{fig3}). Such overluminosity can be explained by an unmodeled effect such as a bright H$\alpha$ emission line from chromospheric activity, interaction with another undetected companion, flaring, accretion or some other unknown process. Assuming that the observed F606W overluminosity is coming from H$\alpha$ emission, we can estimate the strength of emission by calculating the log ratio of its H$\alpha$ luminosity to its bolometric luminosity.  Using the simulated flux normalized spectrum for GQ Lup B, the F606W filter bandpass profile and assuming that all the observed flux in the F606W filter comes from a bright H$\alpha$ emission line, we find GQ Lup B H$\alpha$ emission strength $\log(L_{H\alpha}/L_{\rm{bol}})$ of $-1.71 \pm 0.10$. If this emission if from accretion, it is strong enough to be classified as accreting following the criterion defined in \citet{barrado2003}. This observed emission strength is significantly larger, by an order of magnitude, than what is found for field M, L and T dwarfs \citep{gizis2000} or even in peculiar late type dwarfs \citep{liebert1999,burgasser2000,burgasser2002}. Such peculiar dwarfs are thought to be young low mass objects, $\sim $10 Myr 3-20 M$_{\rm{Jup}}$ \citep{liebert2003}, though not as young as GQ Lup ; GQ Lup B could be a very young example, still bound with its primary, of such objects.

A visible spectrum of GQ Lup B is needed to confirm the H$\alpha$ emission. If the H$\alpha$ emission is present, the emission line 10\% width can be used to discriminate between accretion and chromospheric activity \citep{muzerolle2003,natta2004}. Detection of Pa$\beta$ and Br$\gamma$ lines in the near-infrared could also be used to confirm accretion. Br$\gamma$ was not detected in \citet{neuhauser2005,guenther2005} K-band spectrum, but since this line is harder to detect and fainter than the Pa$\beta$ line, such non-detection does not imply no ongoing accretion \citep{natta2004}. A time series photometric/spectroscopic analysis could distinguish between transient emission due to a strong flare or constant emission characteristic of accretion. Only two R-band images acquired less than four minutes apart are available from the HST archive. The individual photometry of these two images agrees to a few percent, consistent with the constant emission from the accretion hypothesis. Due to the limited time sampling of these observations, more data with bigger time intervals are needed to better quantify the H$\alpha$ flux variability. Searching for an eclipse could also confirm the interacting binary hypothesis. Strong H$\alpha$ emission could also be a sign of a runaway accretion as postulated by \citet{fortney2005} for the core accretion-gas capture model, although such observations would be very fortuitous due to the relatively short timescale expected for this phase. Another more probable possibility is that GQ Lup B simply has form through independent collapse of a molecular cloud fragment at its current separation and is still accreting its circumstellar disk. A PMC search at smaller separations could distinguish the two possibilities. If GQ Lup B is still in its formation phase, it would become a unique object to study ongoing PMC formation.

If the H$\alpha$ emission is from accretion, an unresolved/undetected circumstellar disk must be present around GQ Lup B. Our photometric analysis did not detect an infrared excess at L$^{\prime}$, although the L$^{\prime}$ photometric point is $\sim 1\sigma$ overluminous compared to the models.

\section{Conclusion}
We have reanalyzed available Subaru and HST data to fit GQ Lup B radius and effective temperature using model spectra. Our derived effective temperature ($2335 \pm 100$K) is slightly hotter than the one derived by N05 but with a substantially smaller error bar. Our derived radius for 140pc ($0.38 \pm 0.05$R$_{\odot}$) is more than three times larger than what was found by \citet{neuh2005b} and larger than model predictions. At that distance, our result is consistent to 1$\sigma$ accuracy with a $\sim $15 M$_{\rm{Jup}}$ brown dwarfs (Baraffe models). If GQ Lup B is confirmed to be strongly accreting, it might be a young, still forming/contracting PMC or brown dwarf. A better distance estimate for the GQ Lup system is needed to further constrain GQ Lup B radius and its luminosity. At the one sigma confidence level, our measurement is consistent with a 10-20~M$_{\rm{Jup}}$ companion, at the mass boundary between PMC and brown dwarf. Our results are in agreement with the independent work of \citet{mcelwain2007}.

\acknowledgments
Based in part on data collected at Subaru Telescope, NAO of Japan, and with the NASA/ESA HST, obtained from the STScI data archive (operated by the AURA, NAS 5-26555). This publication makes use of data products from 2MASS, which is a joint project of the University of Massachusetts and the IPAC/Caltech, funded by the NASA and NSF. This research was performed under the auspices of the DOE by the UC, LLNL under contract W-7405-ENG-48, and also supported in part by the NSF Science and Technology CFAO, managed by UCSC under cooperative agreement AST 98-76783. The authors thank Eric Becklin \& Ben Zuckerman for helpful discussions.

\clearpage

\begin{table}
\begin{center}
\caption{GQ Lup B Photometry\label{table1}}
\begin{tabular}{cccccc}\hline
Filter & Date acquired & Vega & mag & mag (Neu05) & Interstellar\\
       &               & zero point & &           & Extinction\\ \hline
F606W & 1999-04-10& 22.92 & $19.19\pm 0.07$ & - & $0.4\pm 0.2$\\
F814W & 1999-04-10& 21.67 & $17.67\pm 0.05$ & - &$0.2\pm 0.1$\\
CH$_4$off& 2002-07-17& -     & $13.76\pm 0.26$ & - &$0.07\pm 0.04$\\
F171M & 2004-04-20& 20.19 & $13.84\pm 0.13$ & - &$0.07\pm 0.04$\\
F190N & 2004-04-20& 18.48 & $14.08\pm 0.20$ & - &$0.06\pm 0.03$\\
F215N & 2004-04-20& 18.25 & $13.40\pm 0.15$ & - &$0.05\pm 0.02$\\
K     & 2002-07-17& -     & $13.37\pm 0.12$ & $13.1\pm 0.1$ &$0.05\pm 0.02$\\
L$^{\prime}$ & 2002-07-17 & -     & $12.44\pm 0.22$ & $11.7\pm 0.3$ &$0.02\pm 0.01$\\ \hline
\end{tabular}
\end{center}
\end{table}

\clearpage

\begin{table}
\begin{center}
\caption{Best Radius and Temperature Fit\label{table2}}
\begin{tabular}{ccccc}\hline
 & Dist. (pc) & IE$_V$ & T$_{\rm{eff}}$(K) & R (R$_{\odot}$)\\ \hline
 &90  & 0.4 & $2335 \pm 80$& $0.25 \pm 0.03$\\
Dist.&140 & 0.4 & $2335 \pm 80$& $0.38 \pm 0.05$\\
 &190 & 0.4 & $2335 \pm 80$& $0.52 \pm 0.07$\\
 & & & & \\
 & 140 & 0.2 & $2320 \pm 80$& $0.38 \pm 0.05$\\
IE &140 & 0.4 & $2335 \pm 80$& $0.38 \pm 0.05$\\
&140 & 0.6 & $2360 \pm 80$& $0.38 \pm 0.05$\\
& & & & \\
Final &140 & 0.4 & $2335 \pm 100$& $0.38 \pm 0.05 d_{\rm{140}}$\\ \hline
\end{tabular}
\end{center}
\end{table}

\clearpage

\clearpage

\clearpage

\begin{thebibliography}{}
\bibitem[Baraffe et al.(2002)]{baraffe2002} Baraffe, I., Chabrier, G., Allard, F., Hauschildt, P.H. 2002, \aap, 382, 563
\bibitem[Baraffe et al.(2003)]{baraffe2003} Baraffe, I., Chabrier, G., Barman, T.~S., Allard, F., Hauschildt, P.~H. 2003, \aap, 402, 701
\bibitem[Barrado et al.(2003)]{barrado2003} Barrado y Navascu\'es, D., Mart\'\i n, E. L. 2003, \aj, 126, 2997
\bibitem[Batalha et al.(2001)]{batalha2001} Batalha, C., Lopes, D.~F., Batalha, N.~M. 2001, \apj, 548, 377
\bibitem[Burgasser et al.(2000)]{burgasser2000} Burgasser, A.~J., Kirkpatrick, J.~D., Reid, I.~N., Liebert, J., Gizis, J.~E., Brown, M.~E. 2000, \aj, 120, 473
\bibitem[Burgasser et al.(2002)]{burgasser2002} Burgasser, A.~J., Liebert, J., Kirkpatrick, J.~D., Gizis, J.~E. 2002, \aj, 123, 2744
\bibitem[Chauvin et al.(2005a)]{chauvin2005a} Chauvin, G. et al. 2005a, \aap, 438, L25
\bibitem[Chauvin et al.(2005b)]{chauvin2005b} Chauvin, G. et al. 2005b, \aap, 438, L29
\bibitem[Fortney et al.(2005)]{fortney2005} Fortney, J.~J., Marley, M.~S., Hubickyj, O., Bodenheimer, P., Lissauer, J.~J. 2005, Astronomische Nachrichten, 326, 925
\bibitem[Gizis et al.(2000)]{gizis2000} Gizis, J.~E., Monet, D.~G., Reid, I.~N., Kirkpatrick, J.~D., Liebert, J., Williams, R.~J. 2000, \aj, 120, 1085
\bibitem[Glass \& Penston(1974)]{glass1974} Glass, I.~S., Penston, M.~V. 1974, \mnras, 167, 237
\bibitem[Guenther et al.(2005)]{guenther2005} Guenther, E. W., Neuh{\"a}user, R., Wuchterl, G., Mugrauer, M., Bedalov, A., Hauschildt, P.~H., Ultralow-mass star formation and evolution, La Palama, June 28, 2005
\bibitem[Hughes et al.(1994)]{hughes1994} Hughes, J., Hartigan, P., Krautter, J., Kelemen, J. 1994, \aj, 108, 1071
\bibitem[Knude \& Hog(1998)]{knude1998} Knude, J., Hog, E. 1998, \aap, 338, 897
\bibitem[K\"onig et al.(2002)]{koenig2002} K\"onig, B., Fuhrmann, K., Neuh\"auser, R., Charbonneau, D., Jayawardhana, R. 2002, \apj, 394, L43
\bibitem[Krist(1993)]{krist1993} Krist, J. 1993, Astronomical Data Analysis Software and Systems II, A.S.P. Conference Series, 52, 536
\bibitem[Liebert et al.(1999)]{liebert1999} Liebert, J., Kirkpatrick, J.~D., Reid, I.~N., Fisher, M.~D. 1999, \apj, 519, 345
\bibitem[Liebert et al.(2003)]{liebert2003} Liebert, J., Kirkpatrick, J.~D., Cruz, K.~L., Reid, I.~N., Burgasser, A., Tinney, C.~G., Gizis, J.~E. 2003, \aj, 125, 343
\bibitem[Mart\'\i n et al.(2000)]{martin2000} Mart\'\i n, E. L. et al. 2000, \apj, 543, 299
\bibitem[McElwain et al.(2007)]{mcelwain2007} McElwain M.~W. et al. ApJ, in press
\bibitem[Mohanty et al.(2004)]{mohanty2004} Mohanty, S., Jayawardhana, R., Basri, G. 2004, \apj, 609, 885
\bibitem[Murakawa et al.(2004)]{murakawa2004} Murakawa, K. et al. 2004, \pasj, 56, 509
\bibitem[Muzerolle et al.(2003)]{muzerolle2003} Muzerolle, J., Hillenbrand, L., Calvet, N., Brice{\~n}o, C., Hartmann, L. 2003, \apj, 592, 266
\bibitem[Natta et al.(2004)]{natta2004} Natta, A., Testi, L., Muzerolle, J., Randich, S., Comer{\'o}n, F., Persi, P. 2004, \aap, 424, 603
\bibitem[Neuh{\"a}user \& Brandner(1998)]{neuhauser1998} Neuh{\"a}user, R., Brandner, W. 1998, \aap, 330, L29
\bibitem[Neuh\"{a}user et al.(2005)]{neuhauser2005} Neuh{\"a}user, R., Guenther, E.~W., Wuchterl, G. Mugrauer, M., Bedalov, A., Hauschildt, P.~H. 2005, \aap, 435, L13
\bibitem[Neuh\"{a}user(2005)]{neuh2005b} Neuh{\"a}user, R. 2005, ESO Workshop Proceedings on Multiple Stars, in press
\bibitem[Reiners et al.(2005)]{reiners2005} Reiners, A., Basri, G., Mohanty, S. 2005, \apj, 634, 1346
\bibitem[Rieke \& Lebofsky(1985)]{rieke1985} Rieke, G.~H., Lebofsky, M.~J. 1985, \apj, 288, 618
\bibitem[Tachihara et al.(1996)]{tachihara1996} Tachihara, K., Dobashi, K., Mizuno, A., Ogawa, H., Fukui, Y. 1996, \pasj, 48, 489
\bibitem[Stassun et al.(2006)]{stassun2006} Stassun, K. G., Mathieu, R. D., Valenti, J. A. 2006, Nature, 440, 311
\bibitem[Wichmann et al.(1998)]{wichmann1998} Wichmann, R., Bastian, U., Krautter, J., Jankovics, I., Rucinski, S.~M. 1998, \mnras, 301, L39
\bibitem[Whitmore et al.(1999)]{whitmore1999} Whitmore, B., Heyer, I., Casertano, S. 1999, \pasp, 111, 1559
\end{thebibliography}
\end{document}